# Probing the Design Space

## Parallel Versions for Exploratory Programming

Tom Beckmann[a] , Joana Bergsiek[a] , Eva Krebs[a] , Toni Mattis[a] , Stefan Ramson[a] , Martin C. Rinard[b] , and Robert Hischfeld[a]

a   Hasso Plattner Institute, University of Potsdam, Germany
b   CSAIL, Massachusetts Institute of Technology, Cambridge, MA, USA

**Abstract**   Exploratory programming involves open-ended tasks. To evaluate their progress on these, programmers require frequent feedback and means to tell if the feedback they observe is bringing them in the right direction. Collecting, comparing, and sharing feedback is typically done through ad-hoc means: relying on memory to compare outputs, code comments, or manual screenshots. To approach this issue, we designed *Exploriants*: an extension to example-based live programming. *Exploriants* allows programmers to place variation points. It collects outputs captured in probes and presents them in a comparison view that programmers can customize to suit their program domain. We find that the addition of variation points and the comparisons view encourages a structured approach to exploring variations of a program. We demonstrate *Exploriants*' capabilities and applicability in three case studies on image processing, data processing, and game development. Given *Exploriants*, exploratory programmers are given a straightforward means to evaluate their progress and do not have to rely on ad-hoc methods that may introduce errors.

## The Art, Science, and Engineering of Programming

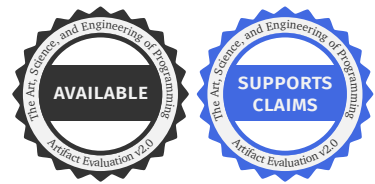

## 1 Introduction

When programmers are confronted with open-ended questions, imprecise requirements, or poorly understood domains, traditional approaches to programming that rely on having well-understood specifications of the desired outcomes fall short. Creative domains and programs with emergent behavior or unknown patterns in their data are particularly unsuited to specify ahead of time.

*Exploratory Programming* is an approach in which programmers write code to prototype or experiment with different ideas for an open-ended goal that may evolve through the process of programming [4]. This approach enables the incremental discovery of solutions along with the desired outcome. Hence, it is common in areas like simulation programming, digital art and music, software engineering, and data science [28].

Exploratory programming is characterized by a trial-and-error nature and programmers' need for backtracking [1, 37] to recover from dead ends or improve on promising experiments. It can be significantly aided by fast iteration cycles and immediate feedback [4, 33]. Practices like ad-hoc versioning and comparing results across versions benefit exploratory programming but are poorly supported by traditional programming environments.

**The Exploratory Ecosystem**   Numerous programming systems and extensions, such as Variolite [17], Juxtapose [12], ForkIt [36], Boba [22], and EditTransactions [24] support backtracking and referring to multiple (micro-)versions by providing ad-hoc versioning in their code editors. Moreover, obtaining feedback is especially convenient in *live programming environments* [28], such as Smalltalk [11], that maintain the impression of changing a running program, and in notebook-style programming environments, like Jupyter [20], that interleave incremental evaluation with intermediate results.

**The Role of Examples**   In large programs, obtaining fast and frequent feedback can still be challenging as programmers need to steer program execution reproducibly to the point where a change in behavior is expected and find ways to inspect (intermediate) output through tools like debuggers or "printf". *Example-based live programming* describes a category of systems that always keep concrete examples close to the edited code so that the consequences of each change emerge immediately. *Babylonian programming* [26] is such a paradigm in which programmers annotate methods with *examples* to create new execution entry points closer to the relevant parts than the main entry point of the program. Programmers can subsequently annotate expressions using *probes* to see the results of the expressions' evaluation. Probes are designed for in-situ, fine-grained feedback inside a code editor.

**Problem Statement and Opportunity**   While immediate feedback on parts of a larger system is currently well supported by example-based programming environments, they lack ad-hoc versioning mechanisms and thus capabilities to back-track and compare behaviors effectively. The latter have been implemented in environments targeting limited scopes, such as notebooks and data analysis pipelines.

Acknowledging that example-based live programming and ad-hoc versioning both contribute positively to the exploratory programming experience, we argue that their combination bridges this gap and can offer fine-grained in-situ feedback across program variations, which is especially promising in larger systems that do not lend themselves to frequent re-runs as a whole.

**Microversioning in Example-based Live Programming**   In this work, we extend the concept of Babylonian Programming with *variation points* in a system called *Exploriants* that brings the microversioning workflow to Babylonian Programming, allowing for effective use of microversioning within large programs while keeping feedback loops short. The best way to display the output is highly dependent on the task at hand – sometimes a simple side-by-side rendering of print strings may suffice, whereas in other cases a domain-specific diffing algorithm offers the most effective insights. Using Babylonian Programming's probes and our variation points, we can collect outputs of interest to the programmer across *all variations of the program*. The collected outputs can then be compared in a customizable user interface. Our reference implementation includes built-in customizations but is designed to be adapted to the user's specific domain.

**Contributions**   As part of this paper, we contribute

- an extension to the Babylonian Programming concept for in-place microversioning and customizable in-situ comparison,
- a demonstration of the practicability of our concept by extending a structured editor in a live programming environment, and
- three exemplary workflows that evaluate *Exploriants*' suitability to meet programmers' requirements throughout an exploratory programming task.

In the following, we first discuss our background (Section 2), followed by a description of the conceptual design of *Exploriants* (Section 3) and its implementation (Section 4). We then describe our exemplary workflows (Section 5). Having established how *Exploriants* can be used, we compare it to related work (Section 6) and subsequently discuss how it addresses the challenges specific to exploratory programming (Section 7), before concluding the paper (Section 8).

## 2   Example-Based Live Programming

In the following, we describe example-based live programming as the paradigm we extend in this work along with prior tools and concepts we use to implement *Exploriants*.

**Live Programming**   Live programming is a way of programming that is characterized by immediate feedback, i.e., the consequences of a change can be observed without disruptive delays, spatially close, and ideally on the same program state that the programmer has been observing before their change without the need to re-evaluate code or re-exercise programs to the point where they show the behavior of interest. Live programming environments, such as Smalltalk [11], support this experience by maintaining the impression of editing the running program [28]. Using the full program with real or manual input, however, limits programmers to a single scenario at a time, and the (concrete) consequences of a change are spatially disconnected from the (abstract) code that caused them.

**Examples**    With examples, programmers can see and understand abstract code together with concrete values. In live programming environments, this can provide users with instant feedback through examples that respond immediately and reproducibly to changes. One of the earliest instances of example-based live programming is *Example Centric Programming* [9], where code is displayed on one panel in the IDE while another panel displays example executions with concrete values. Exemplars [6] facilitate the specification of example inputs for each method in the Newspeak programming system. Examples are similar to tests, leading to tools providing access to examples by making tests related to specific methods easily accessible [30].

**Babylonian Programming**    Babylonian Programming [26] is a form of example-based live programming. Developers create *examples* and are then able to interact with the code itself via widgets. Babylonian Programming features three core concepts:

**Examples**  are live objects (or set-up code creating them) that enable a code region, such as a method, to execute. In object-oriented programming, they need to provide an example instance of a class and example arguments for methods.

**Probes**  are widgets within the code editor that can be placed on any expression and provide domain-specific visualizations of the values recorded during the example execution.

**Replacements**  override expressions during example execution to skip computation that is non-deterministic, long-running, blocking (e.g., waiting for input), or has side-effects. They return a fixed user-specified value instead.

Thus, the examples provide feedback as close to the code as possible. Instead of observing how "the one program" is affected by changes, programmers can isolate smaller scopes within a large program to get feedback.

The Babylonian Programming concept has been implemented for several programming languages and IDEs: For JavaScript in Lively4 [26], for all languages supported by the GraalVM in Visual Studio Code [25], for GDScript in the Godot game engine [21], and for Squeak/Smalltalk in Squeak/Smalltalk [11, 15, 27].

**Sandblocks**    The language-independent structured editor *Sandblocks* allows tool-builders to augment individual expressions [3] and features Babylonian Programming concepts. It is implemented in the live programming environment Squeak/Smalltalk.

For *Exploriants*, we decided to use *Sandblocks* to use our extension together with other features provided by the structured editor while working in a live programming environment. However, our approach can also be applied to other languages, as will be demonstrated in Section 4 with example code for JavaScript.

## 3 Exploriants: Integrated Artifact Management

In this section, we will first explain our design rationale and common problematic aspects of exploratory programming. To address these concerns, we propose the design for an IDE extension, *Exploriants*, that supports programmers in their existing exploratory programming workflows and removes friction in their artifact management.

As described in Section 1, *Exploriants* allows programmers to place variation and observation points (probes) on any expression in the code, to then form all permutations of variation points and illustrate the values captured in probes in an overview. We illustrated a conceptual view on *Exploriants* in Figure 1. Following the design rationale, we describe the concept in more detail using a running example.

### 3.1 Design Rationale

There are several ways for developers to work exploratorily. These ways include but are not limited to simply rerunning code, using live programming, using "printf" statements, or using notebook-style IDEs like Jupyter. The design of our tool was based on previous studies on developer behavior in these exploratory settings, exemplified through the studies below.

A study by Kery et al. on researchers writing code exploratorily noted that researchers often use comments to alter and version code [17]. While this allows them to see different results without additional tooling, it has several disadvantages. It is difficult to switch between different versions of code and to keep in mind which versions exist. Similarly, it is difficult to correlate a created artifact with the code it originates from. Developers need to maintain a mental model of how code maps to artifacts. And once the different versions are no longer needed, researchers will need to clean the code manually.

Another study by Hill et al. with developers working on machine learning found that source code version control is not enough [14]. Artifacts like models and resulting figures are not versioned together with the code. In general, the exploratory process lacks documentation of certain parts: Which code version corresponds to which experiment? Which experiments were there? What is a specific experiment supposed to show? For the participants of the study, it was often necessary to understand previous work done by other people. However, the only method available to them to get this knowledge was to talk to the corresponding other developers.

While more expensive, structured approaches exist, they usually involve custom tools that programmers create or adapt specifically for this one task. It may be unclear whether this upfront investment will pay off, especially if programmers are unsure what artifacts will allow them to effectively evaluate their progress, as observed in a previous study [7].

## 3.2 Factors for Microversioning in Example-Based Live Programming

Our goal is to enable programmers to work with microversioning in larger software systems through example-based live programming. For that purpose, we derive a set of factors from the prior work above and relevant aspects of example-based live programming as described in Section 2.

**Origin of Microversions** As users often work opportunistically [7], it should be possible for users to backtrack and access prior versions of code that they did not know yet would be relevant later.

**Entry Points** Feedback from the running program can only be obtained when execution reaches the relevant parts of the program. As such, it should be possible for users to choose entry points for execution, such that relevant code is reached quickly and ideally without manual intervention.

**Temporal Immediacy of Feedback** To maintain the traits of example-based live programming, feedback needs to be available across relevant microversions in an immediate manner.

**Spatial Immediacy of Feedback** Similarly, artifacts produced from execution should be presented in a manner that supports users in immediately drawing conclusions. This may both mean placing artifacts close to the code that produced them and placing related artifacts of different microversions next to one another for simple comparison.

**Types of Feedback** To support an effective comparison, artifacts should be presented in a rich form that supports users in identifying differences between versions. Ideally, users can also customize the presentation according to their current task.

## 3.3 Exploration through Variation Points

Next, we present our approach to integrated artifact management, *Exploriants*, that is designed to fulfill as many of the factors established in Section 3.2 as possible. We summarize the components of *Exploriants* in Figure 1. In their exploration of a problem space, exploratory programmers continuously formulate programs to validate or explore approaches. We call one concrete approach written in executable code a *universe* and the approach currently written in the source code files the *active universe*.

As an example, we follow Mallory, who is testing if a custom hash set implementation might yield better performance for her workloads than the system hash set implementation she is currently using. Throughout her worfklow, she will interact with the components of Exploriants: *variation points* that consists of a number of *alternatives* to vary expressions in her code, *probes* for marking expressions for feedback, the *grid view* for comparing the results of probes across universes, and the *history view* for recovering variation points from her editing history. An illustration of her workflow is shown in Figure 2.

*Exploriants* reifies the notion of moving from one approach, or universe, to another through *variation points*. Variation points are a means to explicitly fork a universe. In this instance, Mallory selects the expression that constructs the system hash set

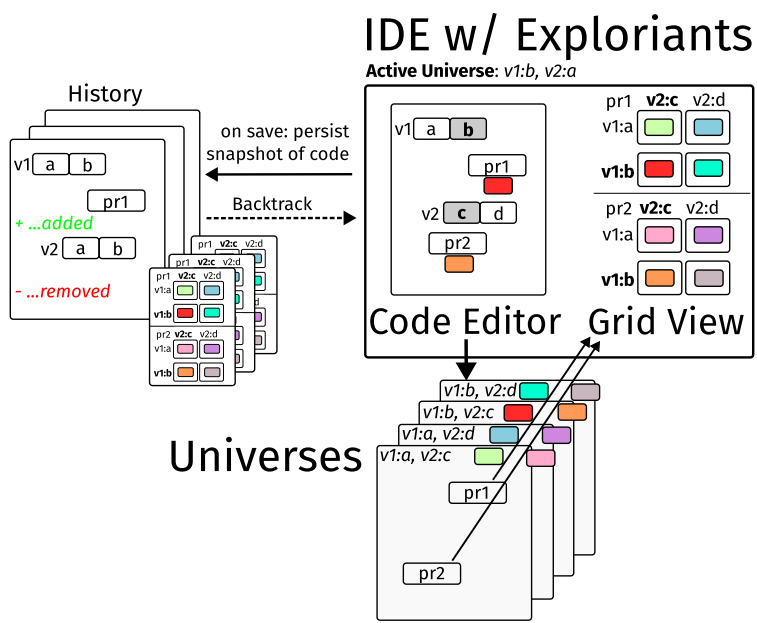

■ **Figure 1** An illustration of an IDE with the *Explorants* extension: the code editor features *variation points* v1 with the two alternatives a and b, and v2 with alternatives c and d. *Probes* pr1 and pr2 report captured values (colored rectangles) for each *universe* to *Explorants*. The captured values for the *active universe,* as configured in each variation point, are shown directly in the code editor. The grid view shows the outputs of probes across all universes. The history allows programmers to recover versions of code they have not persisted in an explicit variation point yet. Snapshots in the history contain the code and the grid view this version of the code produced.

and invokes a shortcut to enclose it in a variation point. The new variation point contains two *alternatives,* both containing her existing code. She now proceeds to modify the second alternative to instead construct her custom hash set. Her custom implementation allows specifying the size of the internal array backing the hash set as a parameter of the constructor. As Mallory is unsure about the right size of the array, she adds a nested variation point for the parameter with three alternatives for an array of sizes 100, 1000, and 10000. *Explorants* now keeps track of four universes, one where the system hash set is used and three where Mallory's custom hash set is used, with a different size parameter each.

At this point, Mallory remembers that she had constructed a test workload earlier in the session before she had decided to try a custom hash set implementation. *Explorants* takes snapshots of the source code on every save in the current session and displays these in a *history* view. If probes are placed, the history view already stores snapshots of the reported objects, allowing programmers to quickly visually identify a prior state of interest. An example of this is shown in Figure 3. Mallory uses the history view to backtrack to the earlier state and copies the relevant source code into a new variation point with alternatives for the specific test workload and the original workload. She now has eight universes, the previous four, alternated by either of the two workloads.

**Figure 2** A sequence of steps illustrating our running example in our reference implementation in Squeak/Smalltalk: (a) Mallory's setup with a linear workload and the system hash set. Mallory selects the expression she wants to vary, and (b) adds a variation point through a shortcut. (c) In the auto-created second alternative, she adapts the expression to create her custom hash set and (d) selects the array size to introduce another variation point. (e) She adds another variation point for the workload, and finally, (f) adds a probe to the expression that produces the execution time. (g) She inspects the execution times in the grid view. Eventually, (h) she uses the cleanup action to remove *Explorants*' extensions from her code.

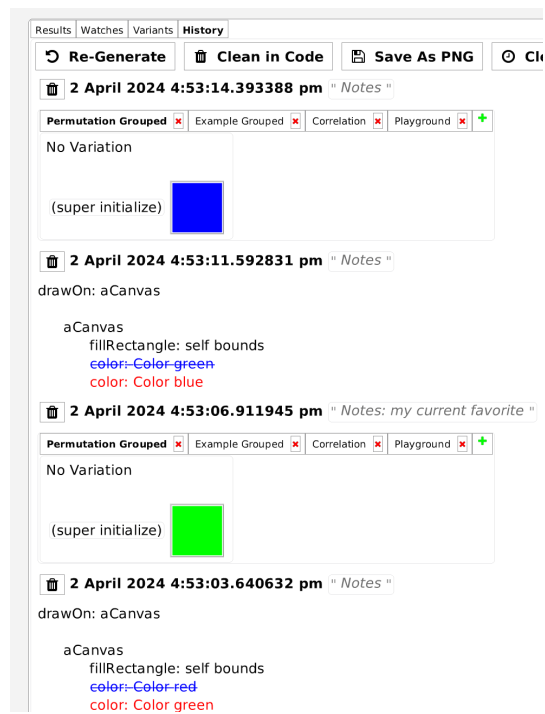

■ **Figure 3** *Explorants*' history view: the programmer change a color value used to draw a rectangle and saved after each change, while a probe was placed on the drawing output. *Explorants* created a snapshot after each save and displays code changes interleaved with code changes.

Variation points and alternatives automatically derive descriptive names from their enclosed source code, as we will describe in Section 4.3. Evidence from prior studies suggests that automatic naming could be important, as programmers during exploratory programming invest little in artifact maintenance [4]. Mallory adapts the auto-generated names for her test and original workload but deems the labels chosen for the constructor and its alternatives, which are shown in Figure 2, adequate to remember what the choices she took meant.

### 3.4 Collecting Artifacts

To get a rough idea of magnitudes, Mallory wants to know how long inserting data from her workloads takes in each universe. She already has code in place that calculates the time spent inserting data. To declare interest in this value, she adds a *probe* around the expression in the code that computes the time via another shortcut. Probes, or watches, are a common means for feedback, especially in live programming contexts [26].

The probe displays the reported value for the *active* universe just below the expression to provide feedback with temporal and spatial immediacy [34]. In this case, Mallory has her custom hash set with an array size of 10000 selected.

The captured values may be any form of data, such as strings, pictures, or composite data structures. The implementation is responsible for defining visualization procedures for common forms of data. The implementation also allows programmers to customize the visualization by defining their own visualization procedure for a form of data of their interest, should built-in visualizations not suffice. In our example, the built-in textual representation for the calculated number works well.

### 3.5 Comparing Universes

To compare values between universes, Mallory could manually switch the active universe and remember the changed values. To simplify this workflow, *Exploriants* not only executes the active universe, but all universes. When this execution encounters a probe, it associates the reported value with the currently executing universe.

Once execution has concluded, all reported values are collected and visualized in the *grid view*, *Exploriants*' user interface for comparing universes and their outputs. Each universe receives a name in the grid view by concatenating the names of variation points and alternatives of its permutation. In essence, the grid view is a detached, side-by-side collection of probes, labeled by the name of the universe that produced the value. The concrete layout of the grid view is configurable. The default layout will collect outputs of each probe in a row and create a new column for each universe. In addition, Mallory configures the workload to appear on the y-axis, as shown in Figure 2, and can now compare a first approximation of the runtime performance for both workloads and her four different hash set configurations.

If a collection of outputs in a certain universe is currently not of interest to programmers, for example, because they want to try an unrelated approach, they can disable an alternative. A disabled alternative persists but does not contribute to the values shown in the grid view. Should an active alternative throw an error, the error is displayed on the user interface, and the alternative is considered disabled for displaying purposes, such that programmers can choose to ignore errors that result from incompatible code changes until the alternative is relevant again.

### 3.6 Sharing and Cleanup

Having concluded her exploratory session, Mallory now wants to share the results with her teammates and persist her findings in case the workload changes and she needs to revisit the decision. *Exploriants* makes use of source code rewriting to persist the location and configuration of variation points and probes, as described in Section 4. Consequently, Mallory only has to create a branch in her version control system and commit the source code as-is, which she can open up again at any later point in an IDE that supports *Exploriants*.

In addition, the grid view supports saving its current state as a picture for direct sharing, e.g., via an instant messaging application. Mallory sends a picture along with a message pointing out what numbers to pay special attention to by referring to the names that *Exploriants* generated for the universes.

Finally, Mallory returns to the main branch, ensures that the active universe reflects the best configuration she found, and uses *Exploriants*' "Cleanup" action. The cleanup removes probes and replaces variation points with the alternative of the active universe, thus removing any traces of her exploratory session in the code. She also adds a comment to the construction place of the hash set, pointing out the name of the version control branch she created, should the exploratory session need to be recovered, in which case her branch could be checked out as-is or merged into the most recent commit.

### 3.7 Customization

There are two types of visualizations in *Exploriants*: *value visualizations*, which are used in the probe and in the grid view to visualize single values, and *grid visualizations*, which receive all universes and all values that were captured during execution of the universes. An implementation of *Exploriants* should at least support visualizing a value by creating a textual representation. Our reference implementation (see Section 4 for more details) also includes built-in visualizations for colors or images.

Grid visualizations facilitate adding automatic ways to compare the reported values. In Mallory's example, she could consider modifying the default grid view in such a way that the fastest, i.e. smallest, run is highlighted in green. Custom grid visualizations for other use cases may include highlighting differences or aligning results in a way that best suits the data.

Users of *Exploriants* can choose the grid visualization through a drop-down. In our reference implementation, code defining new visualizations can be shared among users and be reused across projects.

## 4　Implementation

We built our reference implementation of the *Exploriants* extension for the programming system Squeak/Smalltalk [11, 15] in the Sandblocks IDE [3]. The source code for our reference implementation is published open source.[1] To port *Exploriants* to another IDE, it needs to allow an extension to

- display user interface elements on top of the code,
- nest code editors within those elements, and
- inform the extension when the code is saved.

In the following, we describe relevant aspects of our implementation to enable ports to other environments.

### 4.1 Instrumenting Execution

Our approach generalizes to all programming languages, as we make no specific assumptions about language features. Instead, we use source code rewriting to both mark places where user interface elements need to go and to instrument the source code.

---

[1] The system as described in this paper is available at https://github.com/hpi-swa/sandblocks/tree/64e961a7f8d6b8b5ea3980d11ae954b2229e6e57 (last accessed: 2024-01-20).

Our reference implementation demonstrates the feasibility of our approach for the Smalltalk language. As an example in a different language and to illustrate the instrumentation mechanism, we describe how variations and probes as the two elements of Explorants source code instrumentation can be realized in JavaScript.

**Variations**  Consider the following JavaScript:

```
1  // increment by speed constant
2  position = position + 0.3;
```

Assume the user selects the number 0.3 and invokes the command to wrap it in a variation. The code will then be rewritten in-place to resemble the following, where two identical alternatives using the existing number are automatically created and are ready to be modified by the user:

```
1  // increment by speed constant
2  position = position + __variation({
3      name: '',
4      activeIndex: 2,
5      alternatives: [
6          { name: '', value: () => 0.3, disabled: false },
7          { name: '', value: () => 0.3, disabled: false }
8      ],
9      id: 'e8ufjriu4'
10 });
```

It is then the IDE's responsibility to replace expressions that are marked as variations with a corresponding user interface element. In our reference implementation in Sandblocks, the rewriting occurs in place and displays a user interface as shown in Figure 4. Notably, the Explorants concept would also work without in-place source code rewriting, i.e., when rewriting takes place during a compile step, or without a user interface, i.e., if the above code acts as a user interface.

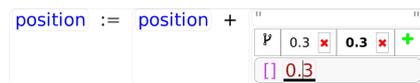

■ **Figure 4**  User interface for varying the speed increase, shown here using our reference implementation for Smalltalk.

An implementation for the __variation marker could look as follows:

```
1  function __variation({ name, activeIndex, alternatives, id }) {
2      // Access a global map of variation ids to alternative indices. While collecting results for all
3      // universes, the map stores which alternatives are meant to be active during the current
4      //   ↪ run.
5      activeIndex = globals.universePermutation[id] ?? activeIndex;
6      return alternatives[activeIndex].value();
7  }
```

The user can indicate the alternative that is running in the default case, i.e. during normal execution of the program, using the `activeIndex` parameter. In the implementation, we first check if we are executing in the context of a specific permutation and otherwise select the user-configured default index. Finally, we execute the closure that corresponds to the selected alternative and return its result.

**Probes**   Probes are implemented using the same rewriting mechanism. Continuing our example, in another place of the source code, the user wants to probe the position variable in the context of a draw call.

```
1  drawRectangle({ x: position, y: 40, width: 10, height: 10 })
```

Upon invoking the command for probing the selected identifier, the code is rewritten to the following:

```
1  drawRectangle({ x: __probe(position, '9af7ih'), y: 40, width: 10, height: 10 })
```

The rewriting step introduces an automatically generated identifier that is used to identify the probe when runtime values arrive. For an IDE that runs in a different process, an implementation of __probe like the following can be used to communicate runtime values:

```
1  function ___probe(value, id) {
2      fetch('http://localhost:9911/report-probe', {
3          method: 'POST',
4          data: JSON.stringify({ id, value })
5      });
6      return value;
7  }
```

The method sends the value in a serialized form to a server on localhost opened by the IDE and then resumes normal control flow by returning the reported value unmodified. The server running inside the IDE would act as follows, where the reported value is passed on to the probe with the reported id:

```
1  server.onPost((request) => {
2      const id = request.body.id;
3      const value = request.body.value;
4      ide.getWatchUserInterfaceForId(id).reportValue(value);
5  });
```

**Implementation Considerations**   Both the JavaScript and our Smalltalk implementations make use of non-evaluated closures to store code that is evaluated by the alternatives only when they are active. An implementation for a language that does not support closures might have to fallback to storing source code in comments that is activated through the source code rewriting step.

As another consideration, the image-based programming system Squeak/Smalltalk does not have a single entry point for program execution [15], so we use Babylonian examples [26] to denote method-level entry points for execution of the universes in Squeak/Smalltalk. When execution of a universe requires user input, our reference implementation supports replacements as defined in Babylonian Programming, to replace evaluation of an interactive statement with a non-interactive expression.

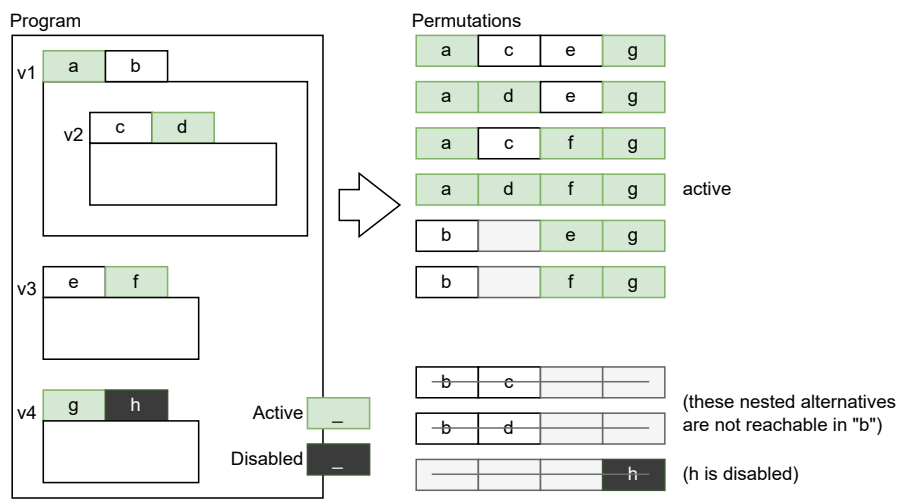

■ **Figure 5**  Illustration of how Exploriants forms permutations, including nested variants. We show a program with four variations, each with two alternatives. Disabled alternatives are marked in dark gray. The active alternative of a variation point is marked green. Disabled alternatives do not contribute to the set of permutations (i.e., "h" is not included). The alternative "a" contains a nested variation "v2", but "b" does not. Correspondingly, "b" is not combined with the nested alternatives. The permutation that is shown in probes, where only a single value is shown, is the permutation marked as active.

## 4.2 Universe Execution

To form all permutations of universes, Exploriants first gathers all variation points that are present in the program. It then forms the Cartesian Product between the alternatives of each variation point, as illustrated in Figure 5. We obtain a set of sequences that each denote a different path through the set of alternatives.

Special care has to be taken when variation points are nested: a nested variation point is only reachable when the alternative it is nested within is chosen in this permutation. Correspondingly, we discard all universes that contain unreachable nested variation points.

Still, the number of universes grows exponentially with each new variation point. From a user's perspective, only a small number of parallel executions will reasonably be helpful, unless the user employs an automated method of comparison. To manage the number of universes, alternatives can be temporarily disabled. Consequently, once users realize that they created too many permutations to be helpful to them, they can disable permutations that appear less promising or relevant, as illustrated in Figure 5, or toggle between disabled sets of alternatives.

### 4.3 Automatically Deriving Names

The name in our reference implementation is derived through three heuristics and updates when the relevant code changes:

- Basic types like numbers, booleans, and strings are named after the textual representation of their value.
- All other expressions show a prefix of the textual form of the expression.
- Variation points consider their calling context: if a variation point's value is used for a named argument in a method call, we name the variation point after the argument; otherwise, we fall back to the first alternative's name.

As an example, given a variation point for the last argument of the expression `image format: 'rgba' kernel: aKernel bias: 3` and the alternatives `1` and `MyBiasConfig darker`, the variation point receives the name `bias:` and the alternatives the names `1` and `darker`, respectively.

These heuristics work best, when alternatives contain small sections of code. To provide a useful name to larger sections, programmers will typically want to manually enter a name.

## 5 Exemplary Workflows

To demonstrate the exploratory programming experience when using *Exploriants*, we identified three exemplary challenges and mapped the associated workflow to *Exploriants*:

1. *Image processing* is a domain where exploration requires frequent visual feedback. We demonstrate the capability to visually review results from image filters with varying parameters and images.
2. *Data processing* with real-world data relies on exploratory discovery and understanding of distributions and outliers. We demonstrate how *Exploriants* helps clean data in a notebook-inspired workflow.
3. *Game development* requires frequent interaction with the game under development to explore and fine-tune the player experience. We demonstrate the novel ability of *Exploriants* to compare and experience *interactions* side-by-side.

### 5.1 Creating New Image Filters

Our first exemplary workflow illustrates the use of the grid view in an open-ended exploration of visual styles and sharing an export of the grid view. Alice is exploring a set of image filters and prepared a test set of two images. She does not have a specific style in mind; instead, her goal is "to have a cool alteration."

Alice creates a new program that loads an image and returns it. She selects the return expression and adds a probe via shortcut. The IDE now shows a preview of the unchanged image in the probe. Next, she adds a variation point for the path of the image and creates alternatives for each of the images in her test set.

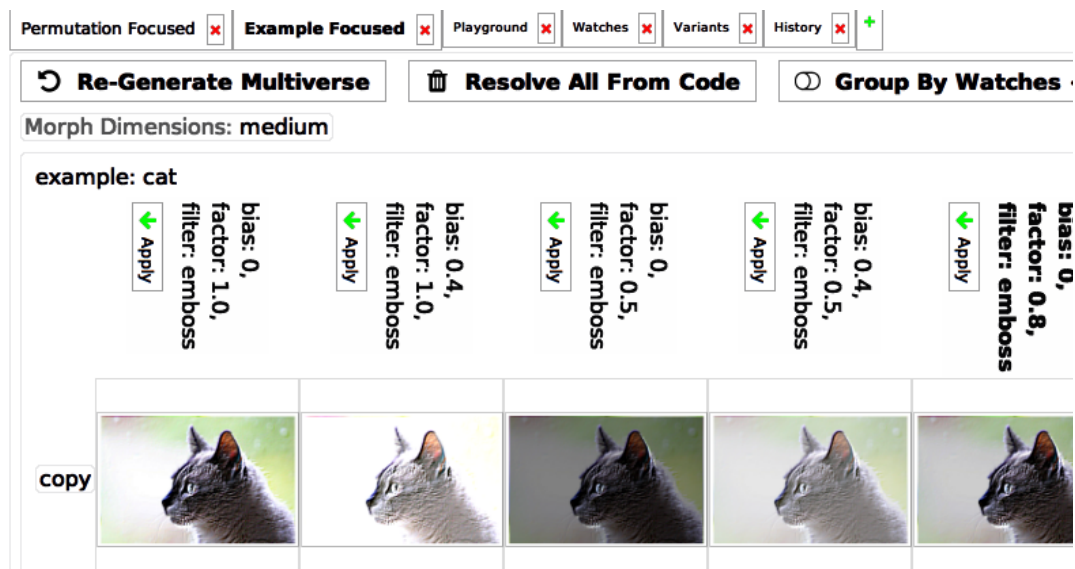

**Figure 6** A custom grid view for viewing multiple versions of a picture at the same time.

Next, Alice adds another variation point to the expression that the method returns. She renames the variation point to "filter" and its first alternative "original". She then proceeds to explore methods in the image class and tries them out, each time adding a new alternative when she finds the result appealing.

Eventually, she discovers the "emboss" method that takes two parameters, "factor" and "bias". To learn the parameters' impact, she changes the values arbitrarily and observes the output in the probe. She now creates a nested variation point for the parameters which are automatically named "factor" and "bias" by *Explorants*, derived from the method signature.

Finally, Alice ends up with three filter variations, one of which is the emboss filter containing three variations for factor and two for bias. She opens the grid view, which presents the probe results for all universes in an overview. She designates that she would like the input image to appear on the y-axis and obtains a grid, an excerpt of which is shown in Figure 6. Each example contains eight columns corresponding to the possible program executions which are labeled by combining all the alternative names that were active in that run.

Alice saves the overview by clicking the "Save as PNG" button and shares it with her peers to present her results and assess which filter is deemed the most popular. Her peers can use the labels above the pictures to refer to the filters. After Alice gathers enough feedback, she clicks "Apply" on the corresponding label, which sets the corresponding universe as active, and uses the resolve button to remove all traces of *Explorants*, leaving just the code for the active universe.

Compared to the other systems discussed in this paper, Explorants facilitates a direct comparison of alternatives in the grid view, as well as an export for sharing purposes. Users also benefit from the first-class, integrated nature of variation points through the clean-up action.

## 5.2 Data Exploration and Cleansing

The second exemplary workflow illustrates how our tool can be used to clean a dataset exploratorily. The objective is to collect source code examples to train an AI model. An important trade-off is data quality vs. quantity, i.e., removing outliers but retaining most of the data.

Bob is tasked with this job. Based on his experience, he expects real-world source code snippets to exhibit a so-called power-law distribution where the vast majority of samples is small (few lines of code), but a few are so large that they make up a significant portion of the data in terms of lines of code and could introduce noise into an AI training procedure.

After loading the dataset, Bob can use it to add an example for the data cleaning method. Bob writes code that bins the data into buckets and plots a histogram in a probe to get an impression of the distribution of code snippet sizes. Unsurprisingly, the distribution has outliers with thousands of lines of code, so he inserts a line that only retains code snippets shorter than a certain cutoff and initializes this variable to an initial guess of 200 LOC to get a readable histogram. The distribution still appears highly skewed toward the left of the histogram, suggesting the threshold was too conservative.

To avoid discarding too much data, Bob inserts a probe that computes the percentage of discarded samples, which initially shows 0.1 %. To adjust the threshold, he adds a variation point on the cutoff variable and tries values 15 and 7. At this point, the editor looks like Figure 7. Using the grid view shown in Figure 8, he can now visually compare the resulting distributions while monitoring the percentage of discarded data and decide whether to continue with the more uniformly distributed but 30 % smaller dataset, or the slightly more skewed distribution that only discards 10 %.

Again, compared to other systems, this workflow demonstrates the possibility for concurrent exploration of alternatives. It also highlights how users can quickly create custom views using the built-in primitives, in this case by adding a probe for both the chart and the percentage value.

## 5.3 Adjusting Player Movement in a Space-Themed Level

This exemplary workflow highlights customization options and sharing an in-progress session. Eve is a game developer, working on a two-dimensional Jump'n'Run game. She aims to adjust the player's movement to feel "light and weightless" when in a space-themed level.

Eve adds a probe on her game object and proceeds to define a custom grid visualization specifically for the returned object type shown in Figure 9. Her custom grid visualization simply renders the given game object but also displays an empty square in the same dimensions as the game view. Inputs in this square are replayed in all other shown game instances.

She then adds variation points for the level's gravity. She starts her cycle of adjusting the gravity values and interacting with the visualization to compare the gravity settings, adding alternatives whenever she finds an interesting permutation. When she adds an

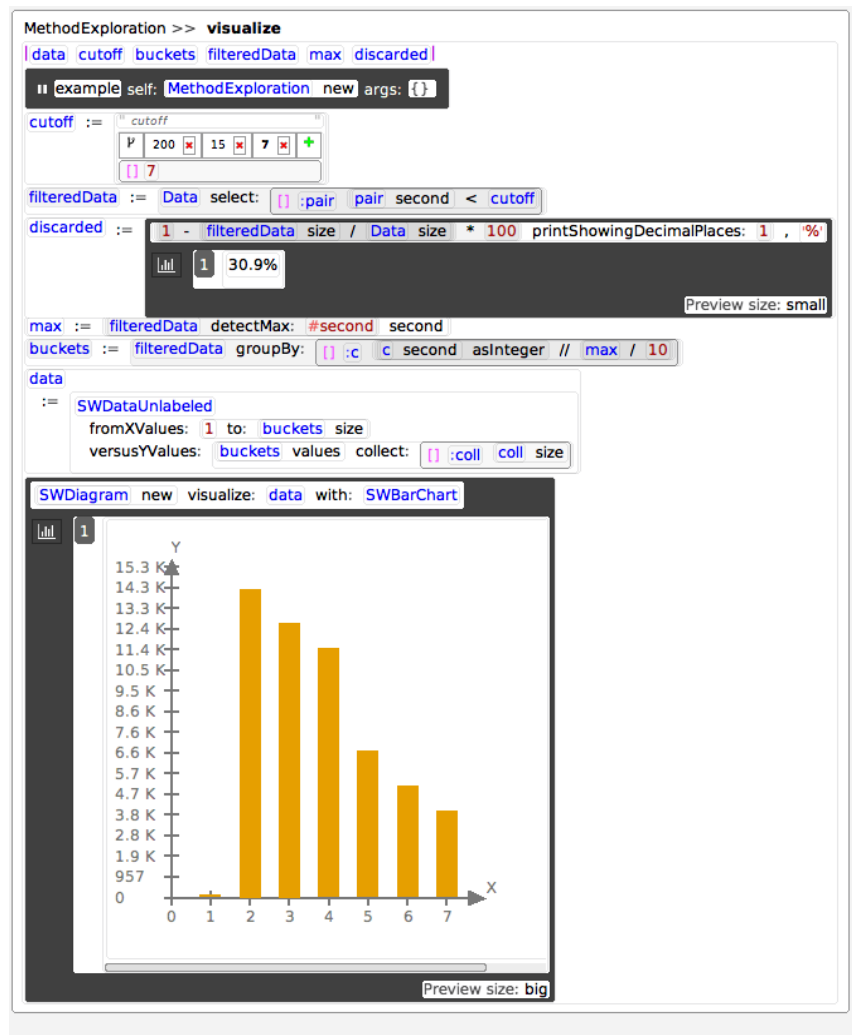

**Figure 7**  A method removing outliers from a dataset and sorting it into buckets. Probes are placed to learn the percentage of discarded data and the final distribution as a histogram. A variation point allows to explore multiple possible cutoff-values.

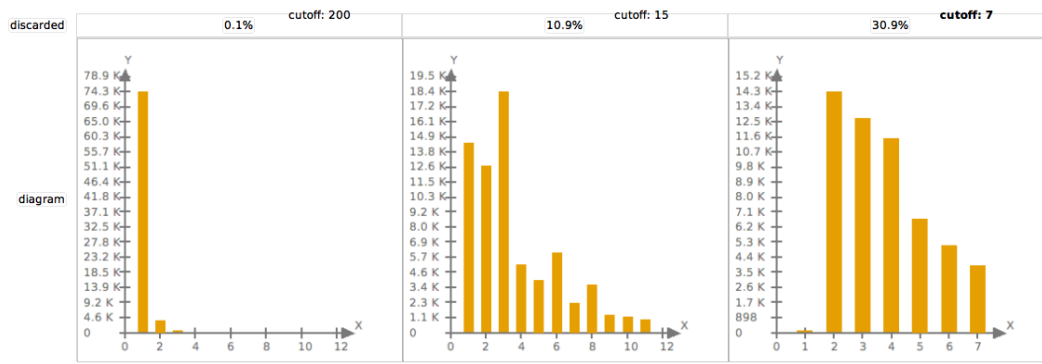

**Figure 8**  A grid view for multiple cutoff values in the data cleaning task in Figure 7. The percentage of discarded samples and the remaining distribution is obtained from the probes and shown side-by-side.

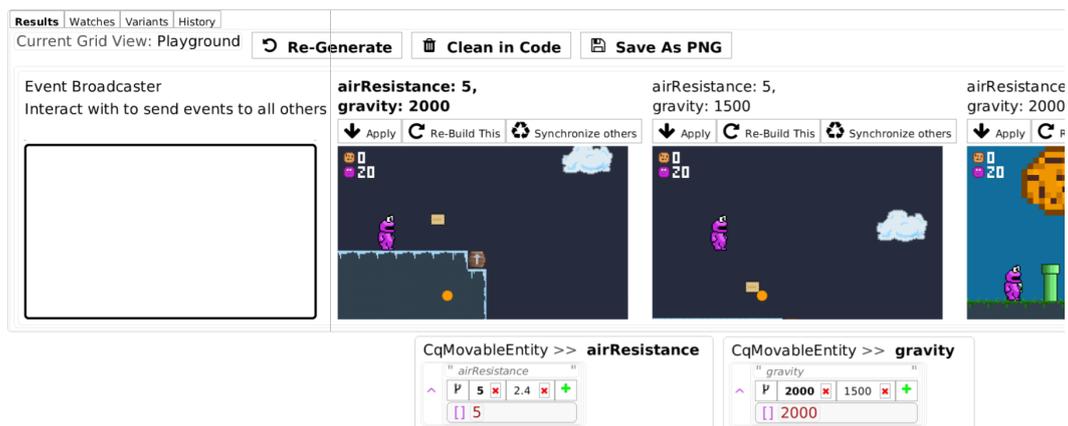

**■ Figure 9** A custom grid view for playing multiple instances of Eve's game at the same time. The empty field on the left has the same dimensions as the game instances and broadcasts events that it receives to all game instances.

alternative, and thus new universes, her custom grid visualization displays the game instances that correspond to the newly defined universes.

To let her teammates try the same, she commits the custom visualization and variation points on a branch, and points her teammates to it. When they open the branch, they are first shown the default visualization in the grid view, which shows the simple textual representation of the game object. They proceed to switch to the custom grid visualization that Eve had created previously to display the game object using the interactive preview and start experimenting as well.

Compared to other systems, this workflow highlights the customizability of the Explorants concept. The same primitives, probes, and variation points, provide the basis for a domain-specific view on the runtime.

## 6 Related Work

In the following section, we review other systems that introduce variation points or artifact comparison to support exploratory programming and compare them to *Explorants* individually. In Section 7, we subsequently discuss how the design of *Explorants* addresses exploratory programming challenges compared to the systems reviewed here.

### 6.1 Variation Points

The idea of a Subjunctive Interface [23] exemplifies the concept that our work adapts specifically for exploratory programming: an interface that supports the user in specifying and subsequently comparing multiple, parallel alternative scenarios. SideViews [31] implements similar concepts, where the outcome of performing a command in a graphical user interface is previewed without performing destructive actions, allowing the user to explore alternatives. In Parallel Paths [32], the user can develop multiple, diverging modifications on an image and compare the outcomes.

In the realm of source code editors, Juxtapose [12] implements an interface where programmers can create copies of source code in alternatives to preview alternative GUI behavior. These alternatives can then be evolved together through linked editing or receive distinct changes. A parameter tuning interface allows for the quick exploration of the impact of changes to variables. Compared to *Exploriants*, Juxtapose does not have a concept of expression-level variation points that form a set of alternative program permutations, as alternatives are defined on a source-file level. Further, *Exploriants* supports dedicated views for arbitrary data, including GUIs.

Variolite [17] adds support for local microversioning on a source line level for exploratory programming tasks. It maintains a branching history of changes when users edit microversions, which is associated with the historical outputs to support backtracking and comparison of outputs across versions. In Variolite, there is only a single active program version at a time, whereas *Exploriants* forms all permutations of alternatives to facilitate comparing alternatives.

Similarly, Fork It [36] extends Jupyter Notebook with support for cell-level alternatives and backtracking to prior versions of cells. Unlike *Exploriants*, alternatives in Fork It cannot be nested; only a single fork may exist at a given time. Fork It does, however, visualize outputs of all forked cells that were executed side-by-side. As a notebook environment, Fork It lends itself to live programming, modifying an evolving program state at run-time by incrementally executing code cells. Notably, it keeps the state for each version separate but synchronizes side effects that affect all versions.

In Boba [22], a tool for multiverse analysis, users can specify alternatives through a custom syntax in a domain-specific language. Boba then acts as a preprocessor that forms all permutations of the defined alternatives and evaluates them. An analysis toolchain allows visual and interactive comparison of model configurations and data, specifically in the data science domain. It is not a live programming environment comparable to *Exploriants* and thus requires a rebuild and restart on every change.

## 6.2 Exploratory Live Programming

Systems can facilitate the comparison of versions and their artifacts by reducing the time it takes to toggle between versions to a minimum, as opposed to showing them side by side. This is in particular facilitated by live programming, where a change to program code often immediately manifests as a change to the visual output of a program. As an example, the time-traveling debugger in Elm [8] allows users to live-reload and even replay prior inputs to their program. Changes to the program can thus be retroactively applied, allowing the user to toggle back and forth between alternative code snippets and evaluate them.

CoExist [29] supports programmers in backtracking during an exploratory programming session, in a method called *explore-first programming*. CoExist allows programmers to go back to a previous code and system state, or even have two states exist in parallel for a comparison. In contrast to Fork It and Boba, which target the data science domain, CoExist focuses on comparing running programs, including their UI. Variation points are not created deliberately as in Variolite, Boba, or Fork It, but recorded during edits. Compared to *Exploriants*, CoExist is centered around a timeline view of versions; individual variations are not visually represented within the currently edited code, but their diff can be viewed by hovering over timeline segments. Live programming allows new changes and backtracking to manifest immediately.

Edit Transactions [24] are a microversioning mechanism specific to live programming environments where code changes would immediately translate into new behavior (e.g., when the active program runs within a simulation or game loop). Instead of taking immediate effect, several atomic changes can be grouped and become effective simultaneously. A notable property is the capability to seamlessly fall back to executing the last working version should the new version fail, allowing programmers to make riskier experiments without restarting the program. Edit Transactions have no dedicated support to compare programs and outputs side-by-side and, similar to CoExist, the versions are not visualized as variation points in the editor.

### 6.3 Observing Artifacts

In exploratory programming, artifacts that manifest as outputs from execution are essential for programmers to evaluate their progress. Various tools and concepts help programmers obtain such outputs faster. Note that these tools are not specifically designed for parallel exploration of alternatives but are still of relevance for artifact collection.

In Subtext [10], source code is merged with its output in the editing interface to provide feedback to programmers close to the source.

Exploratory programming often occurs in the context of data science [4]. Notebooks, such as Jupyter Notebook or Observable, are popular interfaces for data scientists to work with code [19]. In notebooks, multiple statements of code can be grouped in cells, where the output of the last statement of a cell is shown underneath it. Common means to compare outputs in Jupyter Notebook include manually looping over a range and producing multiple outputs, or copying the contents of a cell to produce an alternative. Verdant [13], an extension for Jupyter Notebook, further introduces a history analysis step that allows programmers to understand the evolution of artifacts and facilitates creating a "clean" version of a notebook through archival of experimental computation.

## 7 Discussion

In this section, we will first discuss *Exploriants* in comparison to systems of our related work with regard to our design rationale in Section 3.1 and design factors in Section 3.2. We then analyze *Exploriants* through the Technical Dimensions of Programming Systems [16] to identify design trade-offs and derive avenues for future work. Lastly, we summarize the limitations of our work.

■ **Table 1** An overview of criteria for comparison of exploratory programming systems that relate to Exploriants. We give a description of each point in Section 7.1. Entry points are abbreviated as "Entry P.", and Comparison as "Comp.". A (✓) for immediacy and entry points indicates a notebook.

| System | Variations | | Liveness | | Artifacts | | |
|---|---|---|---|---|---|---|---|
| | Origin | Granularity | Entry P. | Immediacy | Comp. | Positioning | Type |
| Variolite [17] | user, hist. | LOC | | | ✓ | separate | text |
| Juxtapose [12] | user | File | | | ✓ | separate | UI |
| Fork It [36] | user | Cell | (✓) | (✓) | ✓ | between code | text, graph |
| CoExist [29] | history | System | | ✓ | ✓ | separate | program |
| Boba [22] | user | LOC, Expr. | | | ✓ | separate | graph |
| EditTransactions [24] | user | Method | | ✓ | | separate | program |
| Verdant [18] | history | LOC | (✓) | (✓) | ✓ | between code | text, graph |
| Example-centric [9] | | – | ✓ | | | separate | text |
| Babylonian [26] | | – | ✓ | ✓ | | in-line | live objects |
| *Exploriants* | user, hist. | Expr. | ✓ | ✓ | ✓ | in-line, separate | live objects |

## 7.1 Comparing Factors of Systems for Microversioning

Based on our analysis of related work in Section 6 and the factors for microversioning in an example-based live programming workflow we established in Section 3.2, we derived a set of criteria for comparison between *Exploriants* and related systems, as summarized in Table 1.

**Variations** The first set of criteria concerns the *origin of microversions* factor. Half of the systems we analyzed require users to actively decide upfront that they want to introduce a variation point. As users may not always be aware that they would have wanted to keep a version, the other half of systems offers users to go through their edit history to create variations. In Exploriants, the history view is currently lacking dedicated support for automatically recovering past versions, as further described in Section 7.3, but still keeps track of both code changes and the state of artifact output at the time of an older code version.

As additional point of comparison between the systems that emerged during analysis of related systems, we include the granularity of variations. Boba and Exploriants are the only systems that support placing variation points on the lowest level of granularity, expressions. Without expression-level variations, users may need to restructure their code to achieve their desired effect, for example by extracting values to variables such that a line-level variation point can be used. Outside of general-purpose programming, users may benefit from built-in components for structure, such as notebook cells. Here, too, users may have to move code to different cells to achieve a specific effect.

As a third point, most of the systems we analyzed introduce a user interface element inside the editor user interface to access and manage variations. This is in contrast to Boba, where variations are expressed as a fundamental part of the programming language. In CoExist, variations are never expressed explicitly but derived from the history.

**Liveness**   For the second set of criteria, liveness, we first investigate the *entry points* factors. Exampled-based systems, such as *Exploriants*, allow users to finely choose the slice of the program that needs to be executed to obtain feedback by placing an example nearby. In systems that integrate with data science notebooks, users can similarly control the scope of execution by choosing how much code is in each cell but may need to restructure code to accommodate fast entry points. Otherwise, the program will have to be executed from its main entry point, which may delay feedback.

In terms of *temporal immediacy*, some systems we analyzed support live programming, which can significantly lower the delay between change and feedback. This can be implemented by keeping live objects to which changes are migrated or, as is the case in *Exploriants*, by rerunning the relevant part of the program scoped by the example. Other systems are designed around computational notebooks, which may allow users to keep state around to act on after changing their program or choosing a different variation.

**Artifacts**   The third set of criteria concerns artifacts and thus the feedback programmers are presented with. Most of the systems we analyzed have explicit support for comparing artifacts by drawing from outputs of two or more alternative executions and showing them on-screen at the same time. Those that do not, rely on the user to manually keep notes or remember differences between executions.

The placement of artifacts, and thus *spatial immediacy*, presents a challenge to the systems we analyzed: if the artifacts are placed close to the code they relate to, there is only little space available. If they are removed from the code, it may be challenging for users to understand where artifacts are coming from. The latter approach is chosen by most of the systems we analyzed. Notebooks are taking a special role in this regard, as the layout of its code naturally affords placement of large blocks between its cells.

Exploriants' probes attempt to find a middle-ground by showing the output of the active universe in-line with the code but showing the output of all universes in a separate view. While users work to improve the output of the active universe, they can refer to only the local probe. When they have to refer to the full grid of outputs they may be able to more quickly identify the output of relevance by finding the grid that includes the output they just looked at in the probe.

Finally, the *type of the artifacts* the user obtains from the system may significantly constrain their possibilities for an efficient comparison in the general case. Most systems we observed either support forms of presentations that are specific to their target domain, such charts, or just plain text. In Exploriants, users receive the live object graph as observed from the runtime. Using Exploriants' grid view customization, users can then either use the most suitable built-in view on that live object or write snippets of code to customize its display to their use case.

## 7.2 Technical Dimensions of Programming Systems

The Technical Dimensions of Programming Systems (TDPS) [16] suggests a framework for discussing programming systems. It is designed to compare the characteristics of a programming system to others and identify gaps and potentials in a design. In the following, we highlight relevant aspects of the framework based on our evaluation.

*Exploriants*' concept of an active universe that directs the visualization in probes allows it to inherit the fast **feedback loops (TDPS 4.1.19)** from Babylonian Programming [26]. Probes make the impact of a code change visible close to its source, the source code. As its distinguishing characteristic, *Exploriants* allows comparing results between universes via the grid view.

When computations to produce observed values take a while, **immediacy of feedback (TDPS 4.1.2)** may suffer. While baseline performance ultimately always depends on the performance of the code that the user is exploring, it may be possible to fork execution once a variation point is encountered, instead of rerunning the entire program for each permutation. Worlds [35] might demonstrate an approach that facilitates this for some cases.

Creating variation points has two **modes of interaction (TDPS 4.1.4)**, either when explicitly forking a universe through a new alternative, or by recovering a previous, implicit snapshot from *Exploriants*'s history. It would be interesting to investigate unifying these interactions and to always think in terms of historic snapshots. A unified interaction may reduce *Premature Commitment* from the Cognitive Dimensions of Notations (CDN) framework [5], as it no longer require programmers to explicitly introduce alternatives when they might still be unsure if the new direction is worth investigating. On the other hand, without an explicit step to create alternatives, *Visibility* from the CDN may suffer, as transient changes might erroneously be considered variation points.

*Exploriants* supports **expression geography (TDPS 4.2.9)** by lifting overloaded constructs, such as comments, to be distinctive, first-class tools. The variation points and probes offer strong **composability (TDPS 4.3.4)**, as they can nest arbitrarily and act on the expression-level, unlike most prior work. Variation points leave **convenience (TDPS 4.3.5)** to be desired when it comes to dependent variation points: if a single logical alternative requires changes in two disconnected expressions in the code, programmers have to either manually switch two alternatives one after another, or create a variation point that encompasses both. Future work could consequently investigate a means to connect alternatives across variation points.

The concept of the active universe may be considered a trade-off between **convenience** (TDPS 4.3.5) and **conceptual integrity** (TDPS 4.3.2), as well as a trade-off between *Juxtaposability* and *Visibility* of the CDN. The active universe serves to inform inline probes on which universe to display values from. An alternative design could show values from all universes in the inline probe, thus sacrificing *Visibility* as the inline probe would considerably grow in size. If the probe instead was only as a marker and values would be shown only in the grid view, we lose *Juxtaposability* as we spatially disconnect the source, code, from its effect, the probed values.

*Exploriants* allows **addressing customization (TDPS 4.4.2)** points for value and grid visualizations, as described in Section 3.7. Both work in an additive manner where a new visualization can be registered with the system and chosen by the programmer when considered helpful without modifying existing code. Still, future work might need to investigate reducing *Diffuseness* from the CDN: customization may still require too much effort and thus prevent an exploratory programmer from investing time in better tooling. In particular, we hypothesize that programmers might like to remix an existing grid view to quickly adapt domain-specific intricacies in an otherwise well-working layout.

In the **level of automation** (TDPS 4.5.3), generative artificial intelligence could support the ad-hoc workflows we observed in a few ways: naming of alternatives and variation points is already automatic but limited to heuristics that consider local syntactic structures. A future semantic naming scheme might consider the context of the variation point and better summarize the intent of a larger alternative expression. Further, a custom type of grid view could automatically derive and point out semantically meaningful differences in the captured values.

The implementation described in Section 4 makes an effort to provide a high degree of compatibility to existing ecosystems, as discussed in **sociability** (TDPS 4.7.2). Source code rewriting allows integration in any language. Integration of the user interface into the editor may be complicated for some IDEs that either do not support graphical extensions or only allow for narrow extension points. However, the system makes no assumptions about the execution or structure of code, making it compatible with a wide range of editing interfaces, such as structured editors, notebooks, or text editors. In the same dimension of **sociability**, our case studies in Section 5 demonstrate that common workflows from literature can be closely mirrored in *Exploriants* and should thus introduce little resistance for adoption.

Through our own usage of the system and the Technical Dimensions of Programming Systems, we uncovered several important considerations. A future user study could gain further insights that our analysis has missed.

### 7.3 Limitations and Future Work

In the following, we will summarize the most noteworthy limitations of *Exploriants* and briefly discuss potential ways to address them.

**Design Space Explosion**  As described in Section 4.2, the number of variations grows exponentially with the number of variations but users can manually constrain the number by limiting active alternatives to those useful to them at the current time. A future version of Exploriants may also provide automated means to compare, for example, by plotting numbers to show users a large design space condensed into an easily readable format.

**Built-in Custom Views for Comparison**  Exploriants supports customizing the view where users compare outputs of universes. It already has some built-in views that support users in this regard, independent of the domain of the data, for example, by arranging results in a 2D grid by a chosen variation point as shown in Figure 6. Still, users will likely want to have a wider array of built-in comparison methods, such as text diffing, image difference highlighting, or charts.

**Execution Duration and Side Effects**   While execution speed was fast enough to provide live feedback for all workflows described in Section 5, users have to wait for $\sum_i^n t(i)$ seconds, where $n$ is the number of universes and $t(i)$ the time in seconds it takes to execute that universe. To improve execution time, universes could be executed in parallel, leading to a theoretical best case of $max(t(i))$ if the number of cores matches the number of universes. To make parallel execution more efficient, the execution of the program could be modified to create forks at each variation point. In this way, the response time would still be $max(t(i))$ but the time each thread has to be executed is reduced.

Notably, a response time of $max(t(i))$ is inherent to example-based live programming, which we chose as underlying means for execution of Exploriants: the example is always re-run from the start and it is the user's task to choose an example entry point that reduces the amount of code that is executed to a minimum. Similarly, users have to ensure that no side effects can affect parallel or future runs of other universes, such as file accesses. For this purpose, Babylonian Programming, as one instance of example-based live programming, allows users to specify *replacements* for code that triggers side-effects [26].

**Enhanced Backtracking and Sharing**   Exploriants' backtracking functionality does not currently offer a way to share or annotate it. As such, it only serves as backtracking method for the author who is currently exploring a design space. If authors want to show trade-offs they have to manually recover old versions and realize them as variation points. An extended version of the backtracking functionality might allows authors to curate the history and automatically introduce interesting variation points.

## 8   Conclusion

We presented and discussed *Exploriants*, a new concept for an IDE extension for exploratory programming that allows programmers to easily and confidently try out changes to their code and observe and explore all permutations of possible computations based on them. *Exploriants*'s components consists of examples, probes, variations, and a grid view for comparison across universes. We leverage examples from example-based programming to allow for fast feedback on relevant parts of the program, even in large systems.

*Exploriants*'s feedback mechanism integrates with the running system and offers rich customization to support effective comparison tailored to the program domain. We demonstrate capability to work across domains in three exemplary workflows in the domains of data science, visual design, and game design balancing.

**Data Availability**   A virtual machine image including the setup for the example walk-throughs is available at Zenodo [2].

**Acknowledgements** This work is supported by the HPI–MIT "Designing for Sustainability" research program[2] and SAP.

## References

[1] Abdulaziz Alaboudi and Thomas D. LaToza. "Edit - Run Behavior in Programming and Debugging". In: *2021 IEEE Symposium on Visual Languages and Human-Centric Computing (VL/HCC)*. 2021, pages 1–10. DOI: 10.1109/VL/HCC51201.2021.9576170.

[2] Tom Beckmann, Joana Bergsiek, Eva Krebs, Toni Mattis, Stefan Ramson, Martin C. Rinard, and Robert Hirschfeld. *Accepted Artifact for "Probing the Design Space: Parallel Versions for Exploratory Programming"*. Feb. 2025. DOI: 10.5281/zenodo.14717168.

[3] Tom Beckmann, Patrick Rein, Stefan Ramson, Joana Bergsiek, and Robert Hirschfeld. "Structured Editing for All: Deriving Usable Structured Editors from Grammars". In: *Proceedings of the 2023 CHI Conference on Human Factors in Computing Systems, CHI 2023, Hamburg, Germany, April 23-28, 2023*. ACM, 2023, 595:1–595:16. DOI: 10.1145/3544548.3580785.

[4] Mary Beth Kery and Brad A. Myers. "Exploring exploratory programming". In: *2017 IEEE Symposium on Visual Languages and Human-Centric Computing (VL/HCC)*. 2017, pages 25–29. DOI: 10.1109/VLHCC.2017.8103446.

[5] Alan F. Blackwell, Carol Britton, Anna Louise Cox, Thomas R. G. Green, Corin A. Gurr, Gada F. Kadoda, Maria Kutar, Martin Loomes, Chrystopher L. Nehaniv, Marian Petre, Chris Roast, Chris Roe, Allan Wong, and Richard M. Young. "Cognitive Dimensions of Notations: Design Tools for Cognitive Technology". In: *Proceedings of the 4th International Conference on Cognitive Technology: Instruments of Mind*. CT '01. Berlin, Heidelberg: Springer-Verlag, 2001, pages 325–341. ISBN: 978-3-540-42406-2.

[6] Gilad Bracha. *Enhancing Liveness with Exemplars in the Newspeak IDE*. https://newspeaklanguage.org/pubs/newspeak-exemplars.pdf. Accessed: 2025-02-04.

[7] Joel Brandt, Philip J. Guo, Joel Lewenstein, and Scott R. Klemmer. "Opportunistic programming: how rapid ideation and prototyping occur in practice". In: *Proceedings of the 4th International Workshop on End-User Software Engineering*. WEUSE '08. Leipzig, Germany: ACM, 2008, pages 1–5. ISBN: 978-1-60558-034-0. DOI: 10.1145/1370847.1370848.

[8] Evan Czaplicki and Stephen Chong. "Asynchronous functional reactive programming for GUIs". In: *Proceedings of the 34th ACM SIGPLAN Conference on*

*Programming Language Design and Implementation*. PLDI '13. Seattle, Washington, USA: Association for Computing Machinery, 2013, pages 411–422. ISBN: 978-1-4503-2014-6. DOI: 10.1145/2491956.2462161.

[9] Jonathan Edwards. "Example centric programming". In: *ACM SIGPLAN Notices* 39.12 (2004), pages 84–91. DOI: 10.1145/1052883.1052894.

[10] Jonathan Edwards. "Subtext: uncovering the simplicity of programming". In: *ACM SIGPLAN Notices* 40.10 (Oct. 2005), pages 505–518. ISSN: 0362-1340. DOI: 10.1145/1103845.1094851.

[11] Adele Goldberg and David Robson. *Smalltalk-80: The Language and its Implementation*. USA: Addison-Wesley Longman Publishing Co., Inc., 1983. ISBN: 978-0-201-11371-6.

[12] Björn Hartmann, Loren Yu, Abel Allison, Yeonsoo Yang, and Scott R. Klemmer. "Design as exploration: creating interface alternatives through parallel authoring and runtime tuning". In: *Proceedings of the 21st Annual ACM Symposium on User Interface Software and Technology*. UIST '08. Monterey, CA, USA: ACM, 2008, pages 91–100. ISBN: 978-1-59593-975-3. DOI: 10.1145/1449715.1449732.

[13] Andrew Head, Fred Hohman, Titus Barik, Steven M. Drucker, and Robert DeLine. "Managing Messes in Computational Notebooks". In: *Proceedings of the 2019 CHI Conference on Human Factors in Computing Systems*. CHI '19. Glasgow, Scotland Uk: ACM, 2019, pages 1–12. ISBN: 978-1-4503-5970-2. DOI: 10.1145/3290605.3300500.

[14] Charles Hill, Rachel Bellamy, Thomas Erickson, and Margaret Burnett. "Trials and tribulations of developers of intelligent systems: A field study". In: *2016 IEEE Symposium on Visual Languages and Human-Centric Computing (VL/HCC)*. 2016, pages 162–170. DOI: 10.1109/VLHCC.2016.7739680.

[15] Dan Ingalls, Ted Kaehler, John Maloney, Scott Wallace, and Alan Kay. "Back to the future: the story of Squeak, a practical Smalltalk written in itself". In: *Proceedings of the 12th ACM SIGPLAN Conference on Object-Oriented Programming, Systems, Languages, and Applications*. OOPSLA '97. Atlanta, Georgia, USA: ACM, 1997, pages 318–326. ISBN: 978-0-89791-908-1. DOI: 10.1145/263698.263754.

[16] Joel Jakubovic, Jonathan Edwards, and Tomas Petricek. "Technical Dimensions of Programming Systems". In: *The Art, Science, and Engineering of Programming* 7.3 (Feb. 2023). ISSN: 2473-7321. DOI: 10.22152/programming-journal.org/2023/7/13.

[17] Mary Beth Kery, Amber Horvath, and Brad Myers. "Variolite: Supporting Exploratory Programming by Data Scientists". In: *Proceedings of the 2017 CHI Conference on Human Factors in Computing Systems*. CHI '17. Denver, Colorado, USA: ACM, 2017, pages 1265–1276. ISBN: 978-1-4503-4655-9. DOI: 10.1145/3025453.3025626.

[18]  Mary Beth Kery and Brad A. Myers. "Interactions for Untangling Messy History in a Computational Notebook". In: *2018 IEEE Symposium on Visual Languages and Human-Centric Computing (VL/HCC)*. 2018, pages 147–155. DOI: 10.1109/VLHCC.2018.8506576.

[19]  Mary Beth Kery, Marissa Radensky, Mahima Arya, Bonnie E. John, and Brad A. Myers. "The Story in the Notebook: Exploratory Data Science using a Literate Programming Tool". In: *Proceedings of the 2018 CHI Conference on Human Factors in Computing Systems*. CHI '18. ACM, 2018, page 174. DOI: 10.1145/3173574.3173748.

[20]  Thomas Kluyver, Benjamin Ragan-Kelley, Fernando Pérez, Brian Granger, Matthias Bussonnier, Jonathan Frederic, Kyle Kelley, Jessica Hamrick, Jason Grout, Sylvain Corlay, Paul Ivanov, Damián Avila, Safia Abdalla, Carol Willing, and Jupyter development team. "Jupyter Notebooks - a publishing format for reproducible computational workflows". In: *Positioning and Power in Academic Publishing: Players, Agents and Agendas*. Edited by Fernando Loizides and Birgit Scmidt. Netherlands: IOS Press, 2016, pages 87–90. DOI: 10.3233/978-1-61499-649-1-87.

[21]  Eva Krebs, Toni Mattis, Marius Dörbandt, Oliver Schulz, Martin C. Rinard, and Robert Hirschfeld. "Implementing Babylonian/G by Putting Examples into Game Contexts". In: *In Proceedings of the Programming Experience 2024 (PX/24) Workshop, Companion Proceedings of the 8th International Conference on the Art, Science, and Engineering of Programming*. PX/24. Lund, Sweden: ACM, 2024, pages 68–72. ISBN: 979-8-4007-0634-9. DOI: 10.1145/3660829.3660847.

[22]  Yang Liu, Alex Kale, Tim Althoff, and Jeffrey Heer. "Boba: Authoring and Visualizing Multiverse Analyses". In: *IEEE Transactions on Visualization and Computer Graphics* 27.2 (2021), pages 1753–1763. DOI: 10.1109/TVCG.2020.3028985.

[23]  Aran Lunzer and Kasper Hornbæk. "Subjunctive interfaces: Extending applications to support parallel setup, viewing and control of alternative scenarios". In: *ACM Transactions on Computer-Human Interaction* 14.4 (Jan. 2008). ISSN: 1073-0516. DOI: 10.1145/1314683.1314685.

[24]  Toni Mattis, Patrick Rein, and Robert Hirschfeld. "Edit Transactions: Dynamically Scoped Change Sets for Controlled Updates in Live Programming". In: *The Art, Science, and Engineering of Programming* 1.2 (Apr. 2017), Article 13. ISSN: 2473-7321. DOI: 10.22152/programming-journal.org/2017/1/13.

[25]  Fabio Niephaus, Patrick Rein, Jakob Edding, Jonas Hering, Bastian König, Kolya Opahle, Nico Scordialo, and Robert Hirschfeld. "Example-based live programming for everyone: building language-agnostic tools for live programming with LSP and GraalVM". In: *Proceedings of the 2020 ACM SIGPLAN International Symposium on New Ideas, New Paradigms, and Reflections on Programming and Software, Onward! 2020, Virtual, November, 2020*. ACM, 2020, pages 1–17. DOI: 10.1145/3426428.3426919.

[26]  David Rauch, Patrick Rein, Stefan Ramson, Jens Lincke, and Robert Hirschfeld. "Babylonian-Style Programming". In: *The Art, Science, and Engineering of Programming* 3.3 (Feb. 2019), 9:1–9:39. ISSN: 2473-7321. DOI: 10.22152/programming-journal.org/2019/3/9.

[27]  Patrick Rein, Jens Lincke, Stefan Ramson, Toni Mattis, Fabio Niephaus, and Robert Hirschfeld. "Implementing Babylonian/S by Putting Examples Into Contexts". In: *Proceedings of the Workshop on Context-oriented Programming*. COP '19. ACM Press, 2019. DOI: 10.1145/3340671.3343358.

[28]  Patrick Rein, Stefan Ramson, Jens Lincke, Robert Hirschfeld, and Tobias Pape. "Exploratory and Live, Programming and Coding - A Literature Study Comparing Perspectives on Liveness". In: *The Art, Science, and Engineering of Programming* 3.1 (2019), page 1. DOI: 10.22152/programming-journal.org/2019/3/1.

[29]  Bastian Steinert, Damien Cassou, and Robert Hirschfeld. "CoExist: overcoming aversion to change". In: *ACM SIGPLAN Notices* 48.2 (Oct. 2012), pages 107–118. ISSN: 0362-1340. DOI: 10.1145/2480360.2384591.

[30]  Bastian Steinert, Michael Perscheid, Martin Beck, Jens Lincke, and Robert Hirschfeld. "Debugging into Examples". In: *Testing of Software and Communication Systems, 21st IFIP WG 6.1 International Conference, TESTCOM 2009 and 9th International Workshop, FATES 2009, Eindhoven, The Netherlands, November 2-4, 2009. Proceedings*. Volume 5826. Lecture Notes in Computer Science. Springer, 2009, pages 235–240. DOI: 10.1007/978-3-642-05031-2\_18.

[31]  Michael Terry and Elizabeth D. Mynatt. "Side views: persistent, on-demand previews for open-ended tasks". In: *Proceedings of the 15th Annual ACM Symposium on User Interface Software and Technology*. UIST '02. Paris, France: ACM, 2002, pages 71–80. ISBN: 978-1-58113-488-9. DOI: 10.1145/571985.571996.

[32]  Michael Terry, Elizabeth D. Mynatt, Kumiyo Nakakoji, and Yasuhiro Yamamoto. "Variation in element and action: supporting simultaneous development of alternative solutions". In: *Proceedings of the SIGCHI Conference on Human Factors in Computing Systems*. CHI '04. Vienna, Austria: ACM, 2004, pages 711–718. ISBN: 978-1-58113-702-6. DOI: 10.1145/985692.985782.

[33]  Jason Trenouth. "A Survey of Exploratory Software Development". In: *The Computer Journal* 34.2 (Jan. 1991), pages 153–163. ISSN: 0010-4620. DOI: 10.1093/comjnl/34.2.153.

[34]  David Ungar, Henry Lieberman, and Christopher Fry. "Debugging and the experience of immediacy". In: *Commun. ACM* 40 (Apr. 1997), pages 38–43. DOI: 10.1145/248448.248457.

[35]  Alessandro Warth, Yoshiki Ohshima, Ted Kaehler, and Alan Kay. "Worlds: Controlling the Scope of Side Effects". In: *ECOOP 2011 – Object-Oriented Programming*. Edited by Mira Mezini. Berlin, Heidelberg: Springer Berlin Heidelberg, 2011, pages 179–203. ISBN: 978-3-642-22655-7.

[36]  Nathaniel Weinman, Steven M. Drucker, Titus Barik, and Robert DeLine. "Fork It: Supporting Stateful Alternatives in Computational Notebooks". In: *Proceedings of the 2021 CHI Conference on Human Factors in Computing Systems*. CHI '21. , Yokohama, Japan, ACM, 2021. ISBN: 978-1-4503-8096-6. DOI: 10.1145/3411764.3445527.

[37]  Young Seok Yoon and Brad A. Myers. "An exploratory study of backtracking strategies used by developers". In: *5th International Workshop on Co-operative and Human Aspects of Software Engineering (CHASE)*. 2012, pages 138–144. DOI: 10.1109/CHASE.2012.6223012.

## About the authors

**Tom Beckmann** is a member of the Software Architecture Group of the Hasso Plattner Institute at the University of Potsdam. He is working on structured editing for general-purpose languages to better support integration of tools. His current research interests include programming tool design, as well as editing and input methods for programming. Contact Tom at tom.beckmann@hpi.uni-potsdam.de.
https://orcid.org/0000-0003-0015-1717

**Joana Bergsiek** is a graduate student interested in live programming tools. Her research at the Software Architecture Group focuses on quick, intuitive debugging tools for explorative programming. Contact Joana at joana.bergsiek@student.hpi.uni-potsdam.de.
https://orcid.org/0000-0003-4744-0874

**Eva Krebs** is a member of the Software Architecture Group of the Hasso Plattner Institute at the University of Potsdam. Her research interests include example-based programming systems such as Babylonian Programming as well as using live programming systems, gamification, and educational games for computer science education. Contact Eva at eva.krebs@hpi.uni-potsdam.de.
https://orcid.org/0000-0002-9089-7784

**Toni Mattis** is a member of the Software Architecture Group of the Hasso Plattner Institute at the University of Potsdam. His research interests include software modularity and testing, machine learning and AI for programming environments, and code repository mining. Contact Toni at toni.mattis@hpi.uni-potsdam.de
https://orcid.org/0000-0001-7024-9838

**Stefan Ramson** is a member of the Software Architecture Group of the Hasso Plattner Institute at the University of Potsdam. He regards the design of programming systems as the intersection of notation, interface design, psychology, and ergonomics. His current research interests include live and exploratory programming systems, alternative input methods, visual languages, and natural programming. Contact Stefan at stefan.ramson@hpi.uni-potsdam.de.
https://orcid.org/0000-0002-0913-1264

**Martin C. Rinard** is a Professor in the Department of Electrical Engineering and Computer Science at the Massachusetts Institute of Technology and a member of the Computer Science and Artificial Intelligence Laboratory. His research focuses on software systems and related topics, including computer security, program analysis and compilation, machine learning and programming, approximate computing, and software robustness and reliability. Contact Martin at rinard@csail.mit.edu
https://orcid.org/0000-0001-8095-8523

**Robert Hischfeld** leads the Software Architecture Group at the Hasso Plattner Institute at the University of Potsdam. His research interests include dynamic programming languages, development tools, and runtime environments to make live, exploratory programming more approachable. Contact Robert at robert.hirschfeld@uni-potsdam.de.
https://orcid.org/0000-0002-4249-6003